%% file: ms_astro.tex
\documentclass[vecphys]{svmult}

\usepackage{makeidx}         
\usepackage{graphicx}        
\usepackage{multicol}        
\usepackage[bottom]{footmisc}

\makeindex             

\begin{document}

\title*{MAD@VLT: Deep into the Madding Crowd of Omega Centauri}

\titlerunning{MAD@VLT: Deep into the Madding Crowd of $\omega$ Cen} 

\author{G. Bono\inst{1,2} \and A. Calamida\inst{1} \and C. E. Corsi\inst{1}
\and P.B. Stetson\inst{3} \and E. Marchetti\inst{2} \and P. Amico\inst{2} 
\and P.G. Prada Moroni\inst{4} \and I. Ferraro\inst{1} \and G. Iannicola\inst{1} 
\and M. Monelli\inst{5} \and R. Buonanno\inst{6} \and F. Caputo\inst{1}
\and M. Dall'Ora\inst{7} \and S. Degl'Innocenti\inst{4} \and S. D'Odorico\inst{2} 
\and L. M. Freyhammer\inst{8} \and D. Koester\inst{9} \and M. Nonino\inst{10} 
\and A.M. Piersimoni\inst{11} \and L. Pulone\inst{1} \and M. Romaniello\inst{2}
}

\authorrunning{Bono et al.} 

\institute{INAF-OAR, via Frascati 33, 00040 Monte Porzio, Italy 
\texttt{bono@mporzio.astro.it}
\and ESO, Karl-Schwarzschild-Str. 2, 85748 Garching bei Munchen, Germany
\and DAO, HIA-NRC, 5071 West Saanich Road, Victoria BC V9E~2E7, Canada
\and INFN, Sez. Pisa, via Largo B. Pontecorvo 2, 56127 Pisa, Italy
\and IAC, Calle Via Lactea, E38200 La Laguna, Tenerife, Spain
\and Univ. di Roma Tor Vergata, via della Ricerca Scientifica 1, 00133 Rome, Italy
\and INAF-OAC, via Moiariello 16, 80131 Napoli, Italy
\and Centre for Astrophysics, University of Central Lancashire, Preston PR1 2HE
\and Institut Theoretische Physik \& Astrophysik, Univ. of Kiel, 24098 Kiel, Germany
\and INAF-OAT, via G.B. Tiepolo 11, 40131 Trieste, Italy
\and INAF-OACT, via M. Maggini, 64100 Teramo, Italy
}

%
%
\maketitle

\section{Abstract}
We present deep and accurate Near-Infrared (NIR) photometry of the Galactic 
Globular Cluster (GC) $\omega $ Cen. Data were collected using 
the Multi-Conjugate Adaptive Optics Demonstrator (MAD) on VLT 
(ESO). The unprecedented quality of the images provided the 
opportunity to perform accurate photometry in the central 
crowded regions. Preliminary results indicate that the spread 
in age among the different stellar populations in $\omega $ Cen
is limited.    

\section{Introduction}\label{sec:1}

Quantitative constraints concerning the evolutionary properties of 
low-mass stars mainly rely on the comparison between predicted and 
observed Color-Magnitude Diagrams (CMDs) of GCs.
The GCs present several key advantages when compared with 
field stars: {\em i)} cluster stars typically present the same age
and the same chemical composition; {\em ii)} cluster stars are 
located at the same distance, since the depth effects are negligible, 
and present the same reddening; {\em iii)} cluster stars in a CMD 
are distributed according to their evolutionary status, ({\em consecutio}), 
therefore, they are redundant systems; {\em iv)} cluster cores host 
a zoo of compact objects: Cataclysmic Variables (Edmonds et al. 2003), 
Millisecond pulsars (Freire et al. 2003), Low-Mass X-ray Binaries 
(Heinke et al. 2005) and probably either a low or intermediate-mass 
black hole (Bash et al. 2008).     

The main drawback of GCs is that quite often more than half of the 
cluster stars are located in the innermost regions, and indeed
the half mass radius of massive clusters is at most a few arcminutes.
This is the reason why accurate and deep photometry of the innermost 
regions of GCs became chore only with the superb spatial resolution 
and image quality of the Hubble Space Telescope (HST) optical images. 
It is worth mentioning that accurate photometry in the crowded central 
regions of GCs is not a trivial effort even by using HST images. 

Among the Galactic GCs, $\omega$ Cen appears to be a peculiar system. 
It is the only one to show a well defined spread in the abundance of 
iron and $\alpha$-elements, thus suggesting that it might be a 
possible link between GCs and dwarf galaxies. It is also the most massive GC. 
This means that $\omega$ Cen is a perfect laboratory to investigate 
fast evolutionary phases (Calamida et al. 2008) and to constrain the 
chemical enrichment from supernovae ejecta and from previous generations 
of intermediate-mass AGB stars.  
The use of optical and NIR photometry of cluster stars presents 
several advantages when compared with either optical or NIR photometry. 
The stronger temperature sensitivity of optical-NIR colors provides the 
opportunity to select candidate cluster and field stars (Calamida
et al. 2007). For the same reasons, the optical-NIR colors can be adopted 
to pinpoint peculiar stellar populations that present either different 
ages or different chemical compositions (Freyhammer et al. 2005).     
  

\section{Observations and Data Reduction}\label{sec:2}

Optical data were collected with the Advanced Camera for Surveys (ACS) on board 
the HST and the reduction strategy was already  
discussed by Castellani et al. (2007) and by Calamida et al. (2008).

MAD is a prototype instrument performing wide Field of View (FoV) 
real-time correction for atmospheric turbulence (Marchetti et al. 2006). 
MAD has been built by ESO with the contribution of two external consortia 
to prove on the sky the feasibility of Multi-Conjugate Adaptive Optics (MCAO) 
technique in the framework of the 2nd generation VLT instrumentation and of 
the European Extremely Large Telescope (Gilmozzi \& Spyromilio 2007). 
The key advantage of MCAO is to increase the size of
the FoV corrected for atmospheric turbulence. 
The principle of MCAO is based on probing the volume of atmospheric
turbulence above the telescope by performing wavefront sensing on
several guide stars in the FoV. This goal is reached by implementing
a tomographic reconstruction of the turbulence (Ragazzoni, Marchetti
\& Valente 2000) to constrain its 3D distribution and then by applying 
the correction using several deformable mirrors optically conjugated 
at different altitudes in the atmosphere. 

MAD is equipped with three Shack-Hartmann wavefront sensors for measuring
the atmospheric turbulence from three guide stars located in a FoV of
2 arcminutes. Depending on the atmospheric seeing conditions the limiting 
magnitude for the guide stars can be up to V$\sim$13. 
CAMCAO is the MAD NIR camera and it is based on a 2048x2048 pixels Hawaii2 
infrared detector with a pixel scale of 0.028 arcsec per pixel for a total 
FoV of 57.3 arcsec. CAMCAO is mounted on a movable table to scan the 
full 2 arcminutes FoV. The camera is equipped with a standard set of 
$J$, $H$ and $K_s$ filters. MAD has been installed at the Visitor Focus 
of the VLT telescope UT3 (Melipal) located at the ESO Paranal Observatory. 

During the first on-sky demonstration run of MAD two 1x1 arcminute fields
were observed in the region across the center of $\omega$ Cen.
The first field was centered at RA=201.66, DEC=-47.46 and it
was observed on April 3rd, 2007. For wavefront sensing we used three guide
stars of magnitude V$\sim$11.5 and the MCAO loop was closed at a
correction frequency of 400 Hz. Five images of $5\times 24$ sec each were
collected in $Ks$-band and three images of $5\times 24$ sec each in $J$-band.
The second field is centered at RA=201.64, DEC=-47.45 and it was
observed on April 4th, 2007. For wavefront sensing we used three guide stars of
magnitude V$\sim$11.5 equally distributed on a circle of 1 arcmin diameter
concentric to the field and the MCAO loop was closed at a correction
frequency of 400 Hz. Four images of $10\times 24$ sec were collected in
$Ks$-band and three images of $10\times 24$ sec in $J$-band.
The seeing during the observations of the first night changed from 
0.7 to 0.9 arcsec, while during the second it changed from 0.9 
to 1.2 arcsec.

\begin{figure*}
\centering
\includegraphics[height=0.35\textheight,width=0.60\textheight]{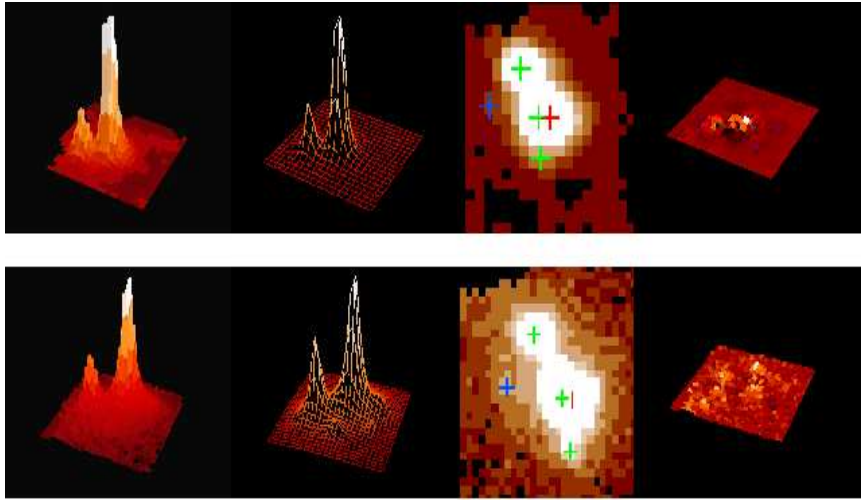}
\caption{Top -- from left to right: the 3D shape of a group of five MS 
stars in a $F435W$ ACS image of 340 sec. Analytical Moffat PSF model of 
the stars plotted in the left panel. Top view of the same stars. The red 
plus marks the brightest star, while the blue plus the star that was 
identified in the MAD image. Residuals after the subtraction of the 
analytical PSF to the data. 
Bottom -- Same as the top panels, but in a 
$K_s$ MAD image of 120 sec. Note the difference in spatial resolution 
of optical and NIR images and the smooth distribution of the 
residuals.     
}
\label{fig:1}       
\end{figure*}

The photometry was performed using DAOPHOT{\small IV}/ALLSTAR and ALLFRAME
(Stetson 1994). We selected $\approx 100$ isolated stars 
to estimate an analytical point-spread function (PSF) for each frame. 
We adopted a Moffat function with $\beta = 2.5$ for the PSF of the $K_s$-band 
images with a quadratic positional change. For the $J$-band images we 
adopted for the PSF either a Moffat, with $\beta = 1.5$, or a Lorentz 
function, and they were assumed linearly variable across the chip. 
We performed PSF analytical photometry on individual NIR images with ALLSTAR. 

Then NIR and optical images were simultaneously reduced using {\sl ALLFRAME}
and the final catalog includes $\approx 7.5\times 10^5$ stars. 
The comparison between the optical and NIR catalogs indicates that 
on MAD images we identified on average more than 90\% ($K_s$-band) 
and more than 75\% ($J$-band) of the stars detected in the same cluster regions
using ACS images.
The photometric calibration of NIR data into the 2MASS system was 
performed using a sample of $\sim\,$ 5000 local standard stars 
(Del Principe et al. 2006).
The accuracy of the absolute zero-point calibration is $\sim\,$ 0.02 mag
for the $K_s$-band and $\sim\,$ 0.03 mag for the $J$-band. We ended up 
with an optical-NIR catalog including $\approx 49,000$ ($K_s$) and 
$\approx 41,000$ ($J$) stars with at least one measurement in an 
optical and in a NIR band.

The images plotted in the upper panels of Fig.~1 show a group of five Main Sequence (MS) 
stars in a $F435W$-band ACS image. The exposure time of this image
is 340 sec and it is located across the cluster center. The brightest 
object in the group is located a couple of magnitudes below the Turn-Off (TO) region
($F435W\,\sim\,19.9$), while the faintest is almost two orders of magnitude
fainter ($F435W\,\sim\,24.5$). The lower panels show the same group of 
stars but in a $K_s$-band MAD image. The exposure time of this image
is 120 sec. The stars quoted above present $K_s$ magnitudes of 
$\sim\,16.9$ and $\sim\,19.4$ mag, respectively. The full-width-half-maximum 
(FWHM) of MAD images is typically better than 0.1 arcsec in the $K_s$-band 
and better than 0.25 arcsec in the $J$-band. This together with the good 
spatial resolution and image quality provided the opportunity to perform 
accurate ground-based PSF photometry in crowded cluster regions. The residuals 
of the fits (rightmost panels) are smaller than 1\% both in the optical and in 
the NIR images. In passing, we note that the faintest star (blue plus) was 
firstly detected in the MAD images and then added in the fit of the ACS 
images.


The NIR bands present several advantages when compared with the 
optical bands. {\em i)} They are marginally affected by reddening 
uncertainties and by the possible occurrence of differential 
reddening (Calamida et al. 2005; van Loon et al. 2007). 
{\em ii)} When moving from MSTO stars down to very-low-mass 
stars the range of NIR magnitudes covered by these structures 
is quite limited $17 \le K_s \le 21$. On the other hand, the same 
structures in the $F435W$ and in the $F625W$ band cover 8 and 7 mag, 
respectively. The difference is due to the 
fact that less massive MS structures are also steadily cooler, therefore, 
their emissivity peaks in the NIR bands.        
    
\begin{figure*}
\centering
\includegraphics[height=0.25\textheight, width=0.85\textwidth]{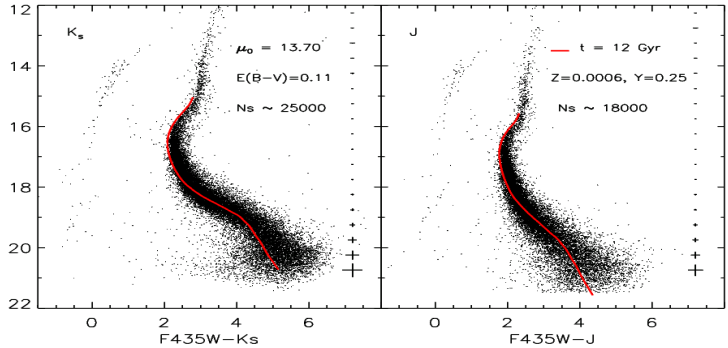}
\caption{Left -- $K_s$ vs $F435W-K_s$ CMD of $\omega$ Cen based on data 
collected with MAD@VLT and with ACS@HST. Current data extend from hot HB stars $K_s\sim 13.5$, 
$F435W-K_s\sim 1.0$ down to the regime of very-low-mass stars $K_s\sim 21$,
$F435W-K_s\sim 6.0$. By assuming $\mu=13.70$, $E(B-V)=0.11$ and evolutionary 
prescriptions by Castellani et al. (2007), in this region of the MS are located 
structures with a stellar mass $M\approx 0.3 M_\odot$. Right -- same as the 
left, but for the $J$ vs $F435W-J$ CMD of $\omega$ Cen.      
}
\label{fig:2}       
\end{figure*}

Data plotted in Fig.~2 show the optical-NIR ($K_s, F435W-K_s$; 
$J, F435W-J$) CMDs of $\omega$ Cen based on MAD and ACS images. 
To our knowledge this is the deepest $K_s, F435W-K_s$ CMD ever collected
for a GC. The fit with a cluster isochrone of 12 Gyr (red line) indicates 
that we detected MS stars with mass values $M \le 0.3 M_\odot$ 
($K_s\approx 21$, left panel). The $J$-band photometry reaches 
similar limiting magnitudes, but the CMD is slightly shallower. 
In passing, we note that current photometry suggests that the 
age spread in $\omega$ Cen appears to be limited, and indeed 
the color age derivative at fixed metal content ($Z=0.0006$) is 
$\Delta (F435W-K_s)/\Delta age \sim 0.06$ mag/Gyr.   

\begin{figure*}
\centering
\includegraphics[height=0.25\textheight, width=0.85\textwidth]{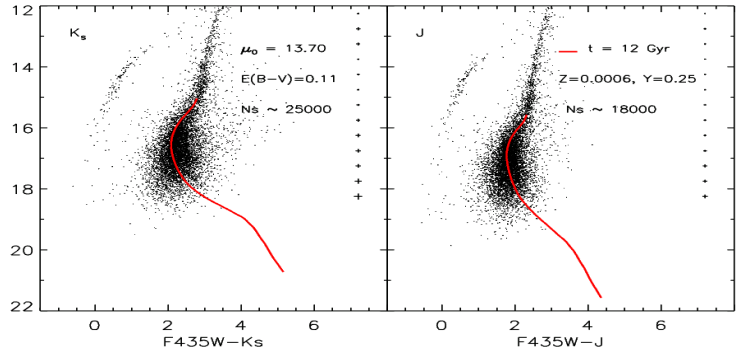}
\caption{Same as Fig.~2, but NIR data collected with SOFI@NTT and 
ISAAC@VLT.}
\label{fig:3}         
\end{figure*}

In order to constrain the impact that MAD has on the photometry of 
crowded regions, Fig.~3 shows the optical-NIR CMD based on data 
of the same cluster regions collected with SOFI@NTT and with 
ISAAC@VLT. The exposure time of 
the NIR images is 240 ($K_s$) and 180 ($J$) sec for SOFI  and 
78 ($K_s$) and 66 ($J$) sec for ISAAC. The observing strategy 
we adopted to reduce these data is very similar to the approach 
discussed in \S 2. The NIR limiting magnitudes of this data set 
is $\approx 3-4$ mag brighter than the MAD data set. The difference 
in the two data sets is mainly due to different seeing conditions: 
0.5--0.7 arcsec  ($K_s$, $J$; SOFI), 
0.4--0.6 arcsec  ($K_s$, $J$; ISAAC) versus 0.1--0.2 arcsec for MAD. 
An important role is also played by the spatial resolution 
$\sim 0.29$ arcsec/px (SOFI), $\sim 0.15$ arcsec/px (ISAAC) versus 
$\sim 0.03$ arcsec/px (MAD). This has a twofold impact on the 
intrinsic accuracy of PSF photometry: {\em i)} the improvement in 
the sampling implies a more accurate image deblending; 
{\em ii)} the decrease in the pixel scale also implies smaller 
fluctuations in the sly background, and in turn, a more efficient 
identification of fainter sources.        

%
%
%
Preliminary results based on NIR data of $\omega$ Cen collected 
with MAD@VLT appear very promising in performing accurate and 
deep photometry in the innermost crowded regions of GCs. Up to 
now these regions have only been investigated using HST. Current 
Adaptive Optics systems have been developed for NIR bands. The 
use of these bands presents several advantages in 
constraining the evolutionary properties of very-low-mass stars 
and the cooling sequence of cluster white dwarfs (Calamida et al. 
2008, in preparation).   

\input{referenc}



\printindex
\end{document}

%% file: referenc.tex
%
%

%
%